\documentclass[graybox]{svmult}
\usepackage{graphicx}
\usepackage{amsmath}
\usepackage{xcolor}
\usepackage{caption}
\usepackage{subcaption}
\usepackage[T1]{fontenc}
\usepackage{float}

\begin{document}

\title*{Accretion Disk-Outflow/Jet and Hard State ULXs}
\author{Mayank Pathak and Banibrata Mukhopadhyay}
\institute{Mayank Pathak \at Joint Astronomy Programme, Department of Physics, Indian Institute of Science, Bengaluru, \email{mayankpathak@iisc.ac.in}
\and Banibrata Mukhopadhyay \at Department of Physics, Indian Institute of Science, Bengaluru \email{bm@iisc.ac.in}}
\maketitle
\vspace{-1.5cm}
\textit{To be published in Astrophysics and Space Science Proceedings, titled "The Relativistic Universe: From Classical to Quantum, Proceedings of the International Symposium on Recent Developments in Relativistic Astrophysics", Gangtok, December 11-13, 2023: to felicitate Prof. Banibrata Mukhopadhyay on his 50th Birth Anniversary", Editors: S Ghosh \& A R Rao, Springer Nature.}
\vspace{0.5cm}

\abstract{Ultraluminous X-ray sources (ULXs) have been objects of great interest for the past few decades due to their unusually high luminosities and spectral properties. A few of these sources exhibit super-Eddington luminosities assuming them to be centering around stellar mass objects, even in their hard state. It has been shown via numerical steady state calculations that ULXs in hard state can be interpreted as highly magnetised advective accretion sources around stellar mass black holes. We use general relativistic magnetohydrodynamic (GRMHD) framework to simulate highly magnetised advective accretion flows around a black hole and show that such systems can indeed produce high luminosities like ULXs. We also verify that the magnetic fields required for such high emissions is around $10^7$ G, in accordance with previous numerical steady state calculations. We further present power profiles for zero angular momentum observer (ZAMO) frame. These profiles show interesting features which can be interpreted as effects of emission due to the Blandford-Znajek and Blandford-Payne mechanisms.}

\section{Introduction}
Ultraluminous X-Ray sources (ULXs) are known to exhibit very high X-ray luminosities. ULXs are point-like, non-nuclear sources which are very rare and most galaxies host either one or none at all. These sources cannot be stars, due to their very high luminosites, which would easily tear apart a star due to radiation pressure. While bright x-ray sources are not uncommon with active galactic nuclei (AGN) having high x-ray luminosities, ULXs are neither as massive as AGNs nor located in galactic nuclei.

Luminosities of ULXs are in the range of $3\times 10^{39} - 3\times 10^{41}$ ergs/s. For a stellar mass source, these luminosities are more than their corresponding Eddington luminosity ($L_{edd}$), which is given by

\begin{equation}
L_{edd}=\frac{4\pi cGM}{\kappa_{es}},
\end{equation}
where $G$ is the gravitation constant, $M$ is the mass of the central gravitating object and $\kappa_{es}$ is the electron scattering opacity.

To explain these high luminosities, ULXs have been modelled as accreting intermediate mass black holes (IMBH) with mass in the range of $10^2-10^4 M_{\odot}$, where $M_{\odot}$ is the solar mass. 
ULXs have also been modelled as stellar mass black holes with slim disks \cite{slim} or radiation pressure dominated super-Eddington accretion flows \cite{sup}.

Another way to explain the super-Eddington emissions from ULXs is by modifying the Eddington limit itself. In the presence of high magnetic fields ($B\geq10^{12} G$), $\kappa_{es}$ reduces, which in-turn reduces the effect of radiation pressure on matter, thereby increasing the effective Eddington luminosity.

A certain number of ULXs also show pulsations which indicates the presence of a neutron star in the system \cite{ns1}\cite{ns2}\cite{ns3}.
However, a few ULXs show power-law spectra in their hard state. This behaviour remains mysterious and quite counter-intuitive. Mondal and Mukhopadhyay (MM19 hereafter) considered numerical steady state calculations to show that this behaviour can be explained by considering ULXs to be highly magnetised sub-Eddington advective accretion flows around a stellar mass black hole \cite{mondal}.

In this paper, we have considered general relativistic magnetohydrodynamic (GRMHD) simulations to explore and verify the model developed in MM19. We have simulated an advective magnetized accretion flow around a stellar mass black hole with various initial conditions and have calculated outflow power from the system. This power comes out be in the range of the observed ULX luminosities. We also discuss outflow power profiles in the zero angular momentum observer (ZAMO) frame. These profiles show interesting behaviour which can be used to determine/separate the Blanford-Znajek and the Blandford-Payne components of the outflow power.
\section{Simulation setup}
We have used the publicly available GRMHD code, BHAC (Black Hole Accretion Code) \cite{bhac} to simulate a system of an accretion disk around a Kerr black hole. The simulation uses horizon-penetrating modified Kerr-Schild (MKS) coordinates to evolve the system and data is output in Boyer-Lindquist (BL) coordinates. We adopt geometric units, i.e., $GM_{BH}=c=1$, $r_{g}=GM_{BH}/c^2=1$ and the light crossing time, $t=GM_{BH}/c^3=1$, in our simulations.
Here, $M_{BM}$ is the mass of the black hole and $c$ is the speed of light.
The accretion disk is initiated by using the Fishbone Moncrief (FM) tours setup \cite{fm}. The black hole spin has been fixed at $a=0.9375$. We have carried out 2.5-dimensional simulations, by exploiting the axisymmetry of the system. The computational domain extends from 1.22 $r_{g}$ to 2500 $r_{g}$ in the radial direction and $0$ to $2\pi$ in the azimuthal direction. The simulations have been run at a resolution of $384\times192\times1$. All simulations are evolved to $3\times10^4$ timesteps. In a future detailed paper, we will elaborate these.

\subsection{Magnetic field evolution}
The simulation solves the induction equation to evolve the magnetic field in curved space-time. The field is initiated by defining the plasma-beta ($\beta$) and initial vector potential. In our simulations, we start with a poloidal magnetic field, initial $\beta=100$ and have used the following vector potentials to initiate SANE (Standard and Normal Evolution) and MAD (Magnetically Arrested Disk) accretions \cite{kc}, respectively:
\begin{enumerate}
    \item $A_{\phi}=\max(\rho/\rho_0-0.2,0)$,
    \item $A_\phi=\exp(-r/r_{o})(r/r_{in})^3\sin^3\theta\max(\rho/\rho_{0}-0.01,0)$,
\end{enumerate}
where $\rho_0$ is maximum density, $r_{in}$ is inner edge of the disk and $r_o=400 r_g$.
Note that GRMHD simulations with same parameters have also been carried out using the alternate publicly available code HARMPI by Raha et al. 2024 \cite{raha}. These results are also published in the present volume.

 \begin{figure}
 \begin{subfigure}[b]{0.47\textwidth}
\includegraphics[width=\textwidth]{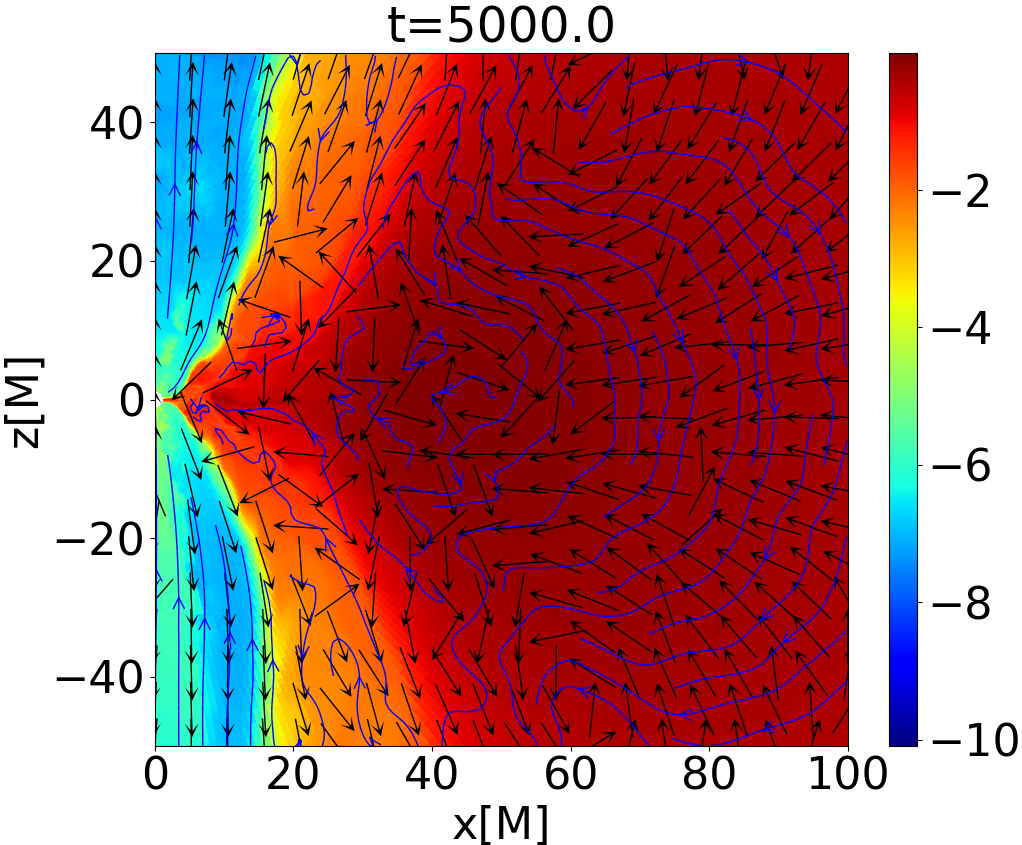}
\end{subfigure}
\begin{subfigure}[b]{0.47\textwidth}
\includegraphics[width=\textwidth]{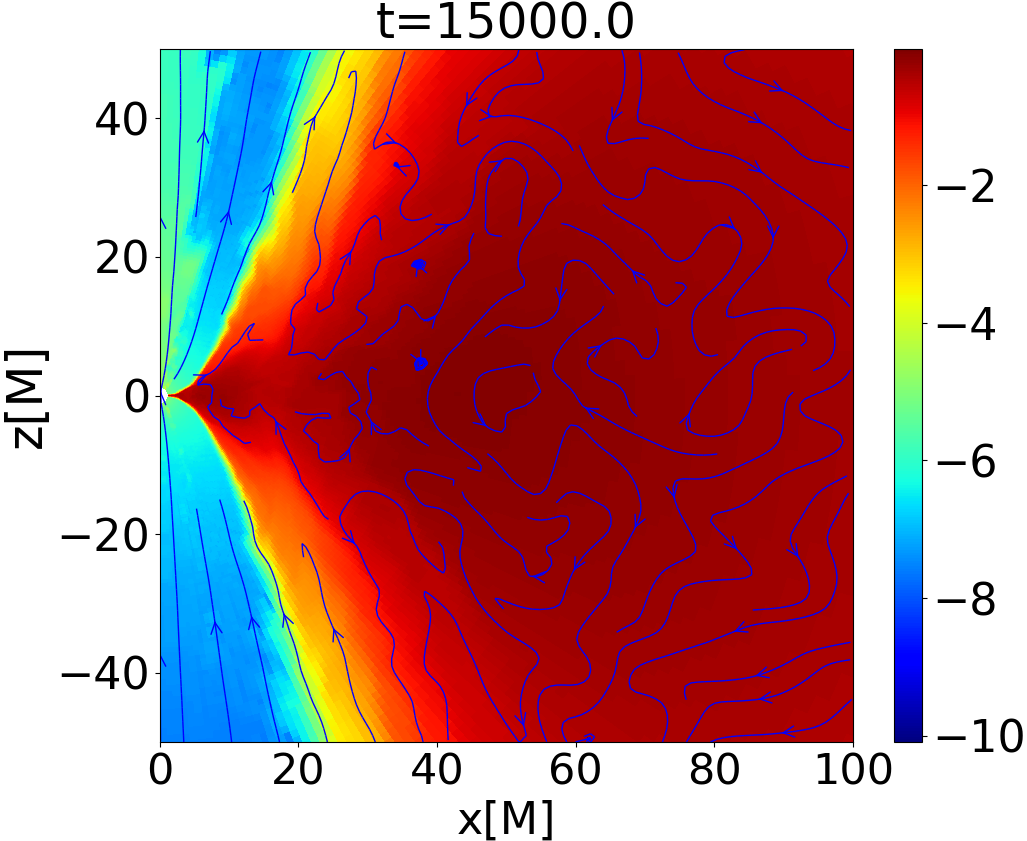}
\end{subfigure}
\begin{subfigure}[b]{0.47\textwidth}
\includegraphics[width=\textwidth]{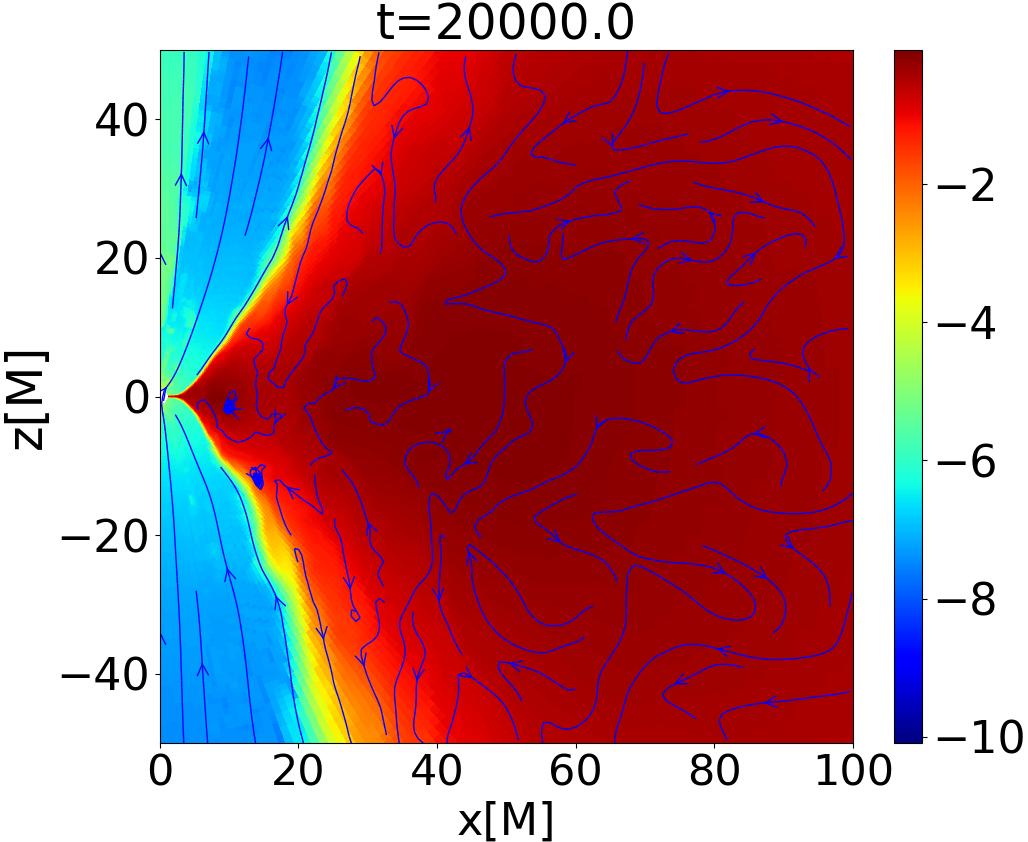}
\end{subfigure}
\hspace{0.18in}
\begin{subfigure}[b]{0.47\textwidth}
\includegraphics[width=\textwidth]{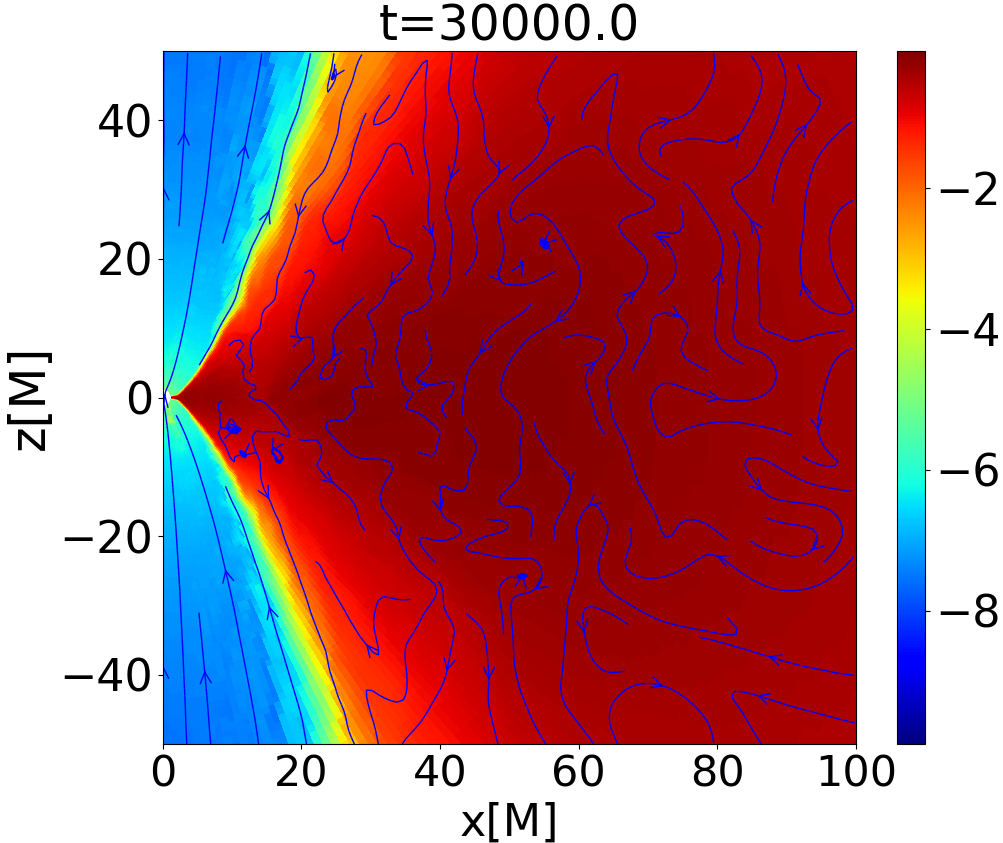}
\end{subfigure}
     \caption{Density contours for SANE with magnetic field streamlines. The top-left panel shows the velocity vectors.}
     \label{sane}
 \end{figure}
 
\section{Results}
\begin{figure}
 \begin{subfigure}[b]{0.48\textwidth}
\includegraphics[width=\textwidth]{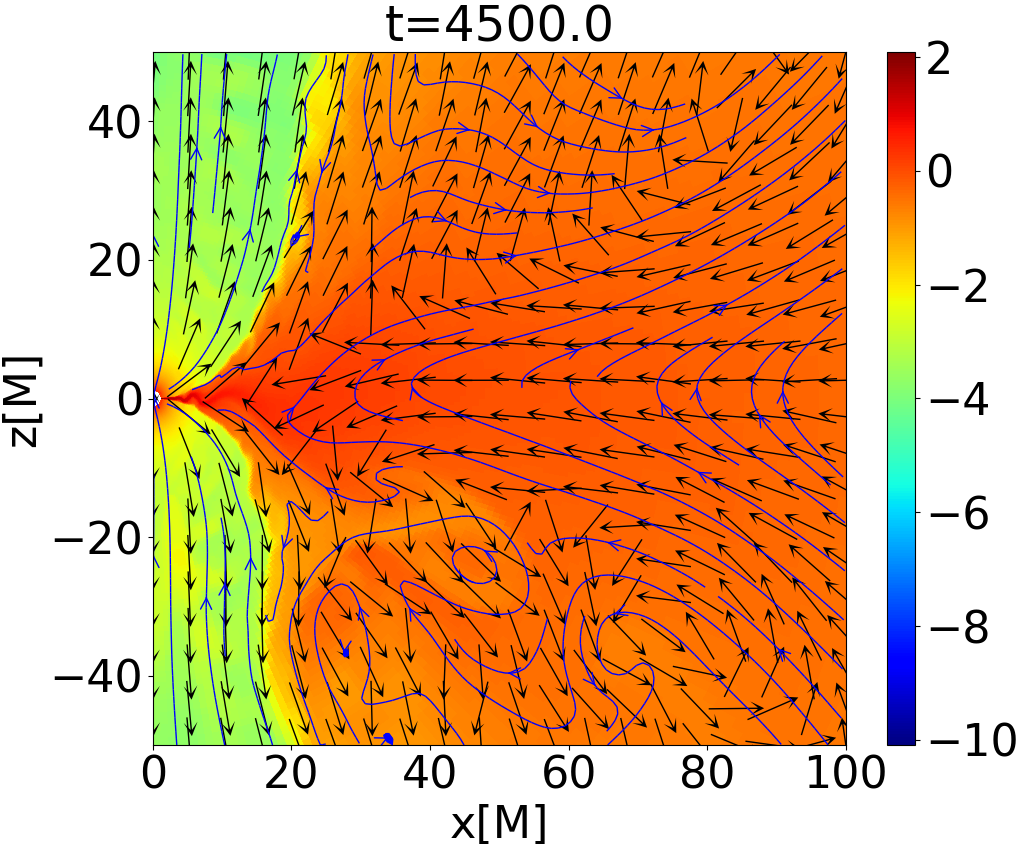}
\end{subfigure}
\begin{subfigure}[b]{0.47\textwidth}
\includegraphics[width=\textwidth]{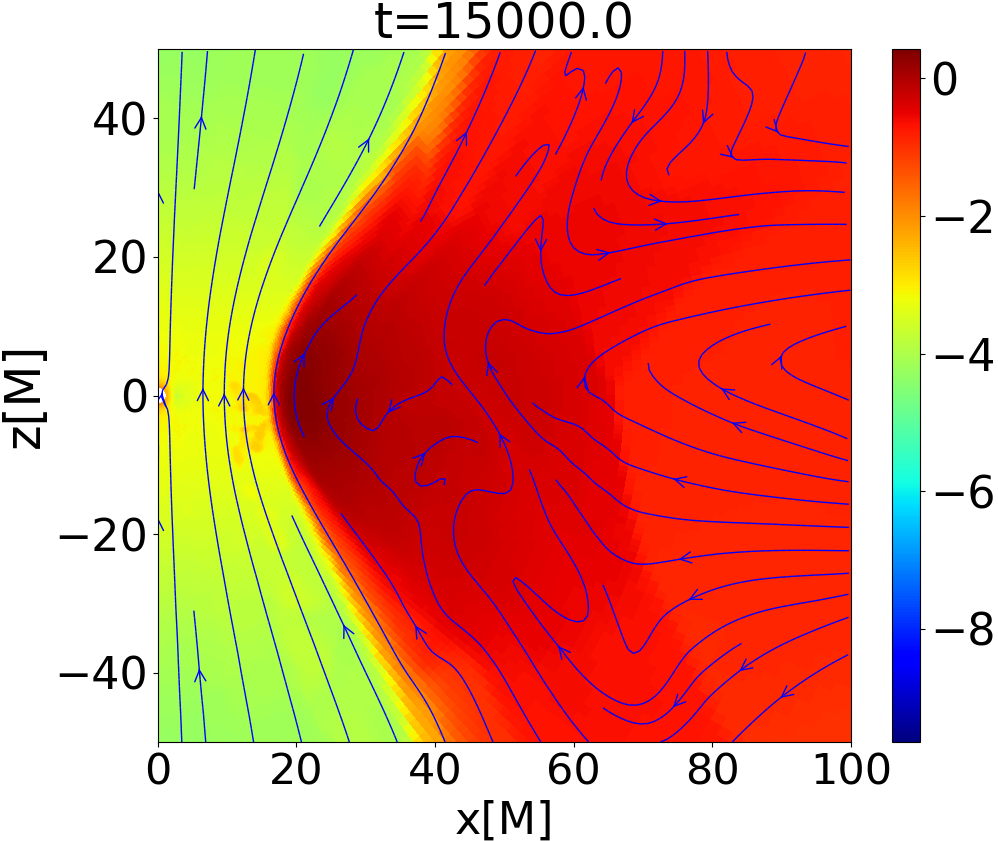}
\end{subfigure}
\begin{subfigure}[b]{0.47\textwidth}
\includegraphics[width=\textwidth]{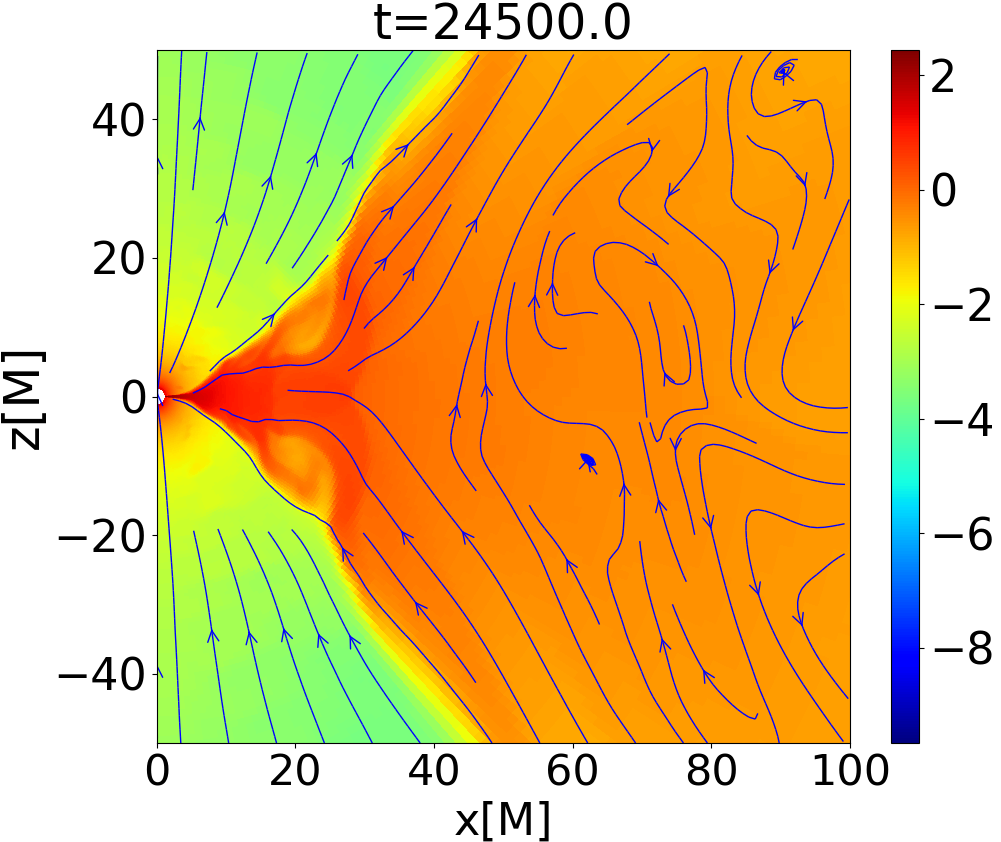}
\end{subfigure}
\hspace{0.18in}
\begin{subfigure}[b]{0.47\textwidth}
\includegraphics[width=\textwidth]{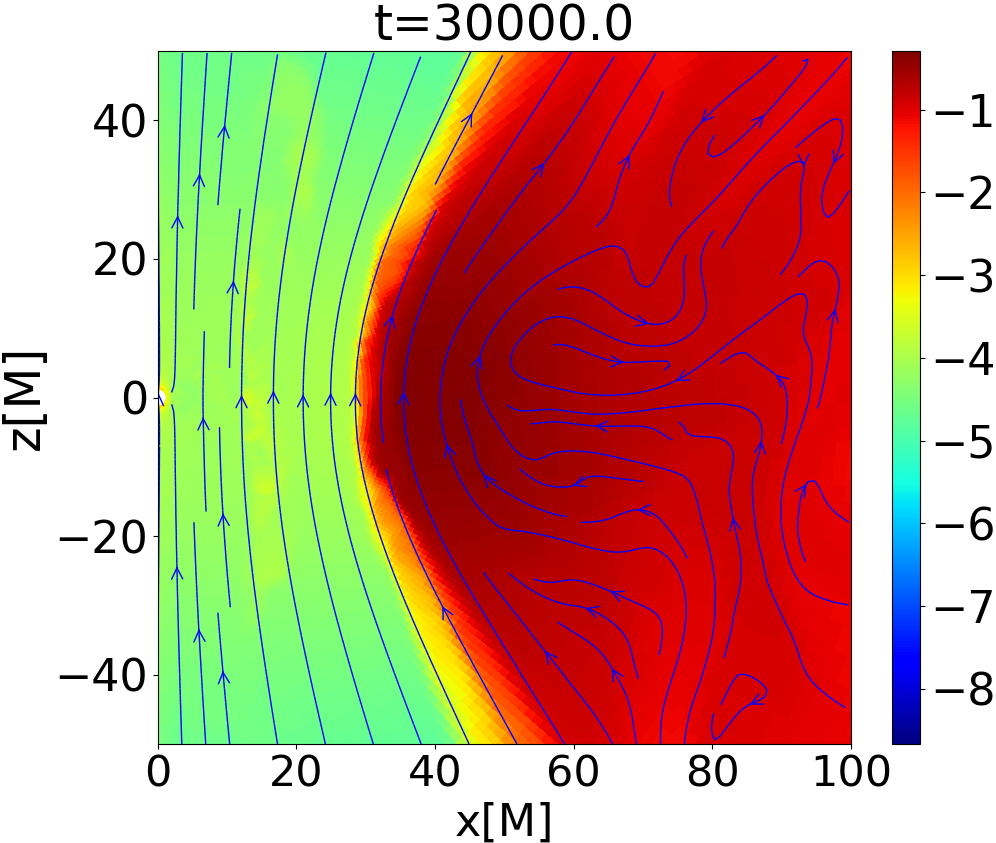}
\end{subfigure}
     \caption{Density contours for MAD with magnetic field streamlines. The top-left panel shows the velocity vectors.}
     \label{mad}
 \end{figure}

Figs. \ref{sane} and \ref{mad} show density contours with magnetic field streamlines at different time for SANE and MAD simulations respectively. These contours show the evolution of flow density and magnetic fields over the course of the simulations.

As evident from Fig. \ref{sane}, the SANE accretion flow is uninterrupted throughout the evolution. At later times, the flow is accreted via a thin funnel onto the black hole. This is due to the formation of strong magnetic fields near the black hole, which oppose the accretion flow via magnetic pressure. 

Fig. \ref{mad} shows that the accretion is not continuous for MAD systems and the flow oscillates due to the formation of a barrier near the black hole. This is because MAD flows accumulate as much magnetic flux as possible in a short evolution time, leading to large magnetic pressure near the black hole which in-turn results in a magnetic barrier. Due to the turbulent nature of the flow, the ram pressure of the accreting matter occasionally breaks through the magnetic barrier but due to high magnetic flux accumulation, the barrier forms again. This effect is exaggerated in 2-dimensional simulations, as the accretion flow is confined to only one azimuthal plane.

To explore the outflow power of the flow, we define the following two quantities:
\begin{enumerate}
    \item Mass Accretion rate ($\Dot{M}$):
\begin{equation}
    \Dot{M}(r)=-\int\sqrt{-g}\rho u^rd\theta d\phi.
\end{equation}
    \item Inward energy flux ($\Dot{E}$):
\begin{equation}
    \Dot{E}(r)=\int \sqrt{-g}T^r_td\theta d\phi,
\end{equation}
\end{enumerate}
where $g$ is the metric determinant and $T^{\mu}_{\nu}$ is the stress energy tensor, given by
\begin{equation}
    T^{\mu}_{\nu}=(\rho+\tilde{P}+\tilde{P}/(\gamma+1)+b^2)u^{\mu}u_{\nu}-b^{\mu}b_{\nu},
\end{equation}
where $\rho$ is the density, $\tilde{P}$ is the pressure of the flow and $\gamma$ is the adiabatic constant; $u^{\mu}$ and $b^{\mu}$ are the four-velocity and four-magnetic field respectively and $b^2=b^{\mu}b_{\mu}$.

Here the signs are chosen such that positive values mean the flow of the quantity into the black hole.
The net output power is then defined as \cite{kc}:
\begin{equation}
    P(r)=\Dot{M}(r)-\Dot{E}(r)=-\int\sqrt{-g}\rho u^rd\theta d\phi-\int \sqrt{-g}T^r_td\theta d\phi \label{power}
\end{equation}
The quantity $P(r)$ calculated following eq. (\ref{power}) gives the net outflow power in code units. To infer the physical power from $P(r)$, we need to multiply it with $\Dot{M}_sc^2$, where  $\Dot{M}_s$ is the scale of accretion rate, defined by $\Dot{M}_{phy}=\Dot{M}_s\Dot{M}$, with $\Dot{M}_{phy}$ being the physical accretion rate. Since we are considering advective accretion flows, we have chosen $\Dot{M}_{phy}=0.05\Dot{M}_{edd}$, where $\Dot{M}_{edd}$ is the Eddington accretion rate, given by, $\Dot{M}_{edd}=1.39\times10^{18}(M_{BH}/M_{\odot})$ gm/s and $M_{BH}=20M_{\odot}$.

Thus, the dimensional power is given by
\begin{equation}
    P(r)=\left(\frac{\Dot{M}(r)-\Dot{E}(r)}{\Dot{M}(r')}\right)\Dot{M}_{phy}c^2 \label{dpower},
\end{equation}
where $r'$ is the radius chosen for normalising the outflow power. 

The definition of time averaged power is ambiguous. Thus, we have considered three power definitions based on the types of time-averaging: 
\begin{enumerate}
    \item $P(r)=<\Dot{M}(r)-\Dot{E}(r)>/<\Dot{M}(r)>\Dot{M}_{phy}c^2$,
    \item $P(r)=<\Dot{M}(r)-\Dot{E}(r)>/<\Dot{M}(r_{eq})>\Dot{M}_{phy}c^2$,
    \item $P(r)=<\Dot{M}(r)-\Dot{E}(r)>/<\Dot{M}(r_h)>\Dot{M}_{phy}c^2$,
\end{enumerate}
where $r_{eq}$ is the steady flow radius and $r_h$ is the event horizon radius.
\subsection{Steady flow radius}
The accretion flow in our simulation domain undergoes inflow and outflow. As the simulation evolves, the flow reaches an inflow-outflow equilibrium out to a certain radius \cite{rn}. This radius is the steady flow radius ($r_{eq}$). Depending on the initial magnetic vector potential and time of evolution, $r_{eq}$ changes.
To determine $r_{eq}$, we investigate the time averaged accretion rate profiles for both SANE and MAD simulations.

 \begin{figure}
 \begin{subfigure}[b]{0.48\textwidth}
\includegraphics[width=\textwidth]{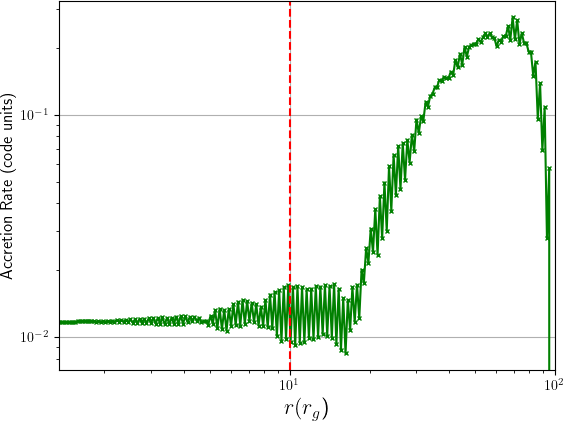}
\caption{SANE}
\end{subfigure}
\begin{subfigure}[b]{0.48\textwidth}
\includegraphics[width=\textwidth]{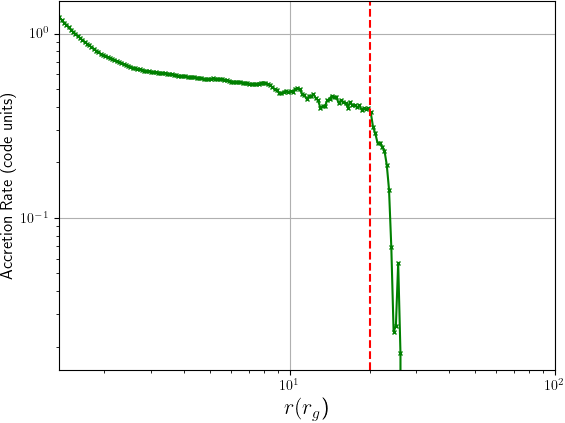}
\caption{MAD}
\end{subfigure}
     \caption{Time averaged accretion rate profiles. Red dashed line indicates $r_{eq}$.}
     \label{acc}
 \end{figure}

As evident from Fig. \ref{acc}, the radii $r_{eq}$-s for the SANE and MAD simulations are 10 $r_g$ and 20 $r_g$ respectively \cite{me}. The lower value of $r_{eq}$ for SANE indicates the slower evolution of the system compared to MAD. This is because MAD systems accumulate much more magnetic flux near the black hole in a smaller amount of time as compared to SANE. Thus MAD systems have more time to equilibrate with the conditions within the simulation domain. We compute the outflow power only from the inflow-outflow equilibrium region of the accretion flow, i.e., from the horizon of the black hole to $r_{eq}$. The region outside $r_{eq}$ has not achieved inflow-outflow equilibrium and is thus unphysical from an astrophysical standpoint.

\subsection{Outflow power profiles}

\begin{figure}
\centering
\begin{subfigure}[b]{0.83\textwidth}
\includegraphics[width=\textwidth]{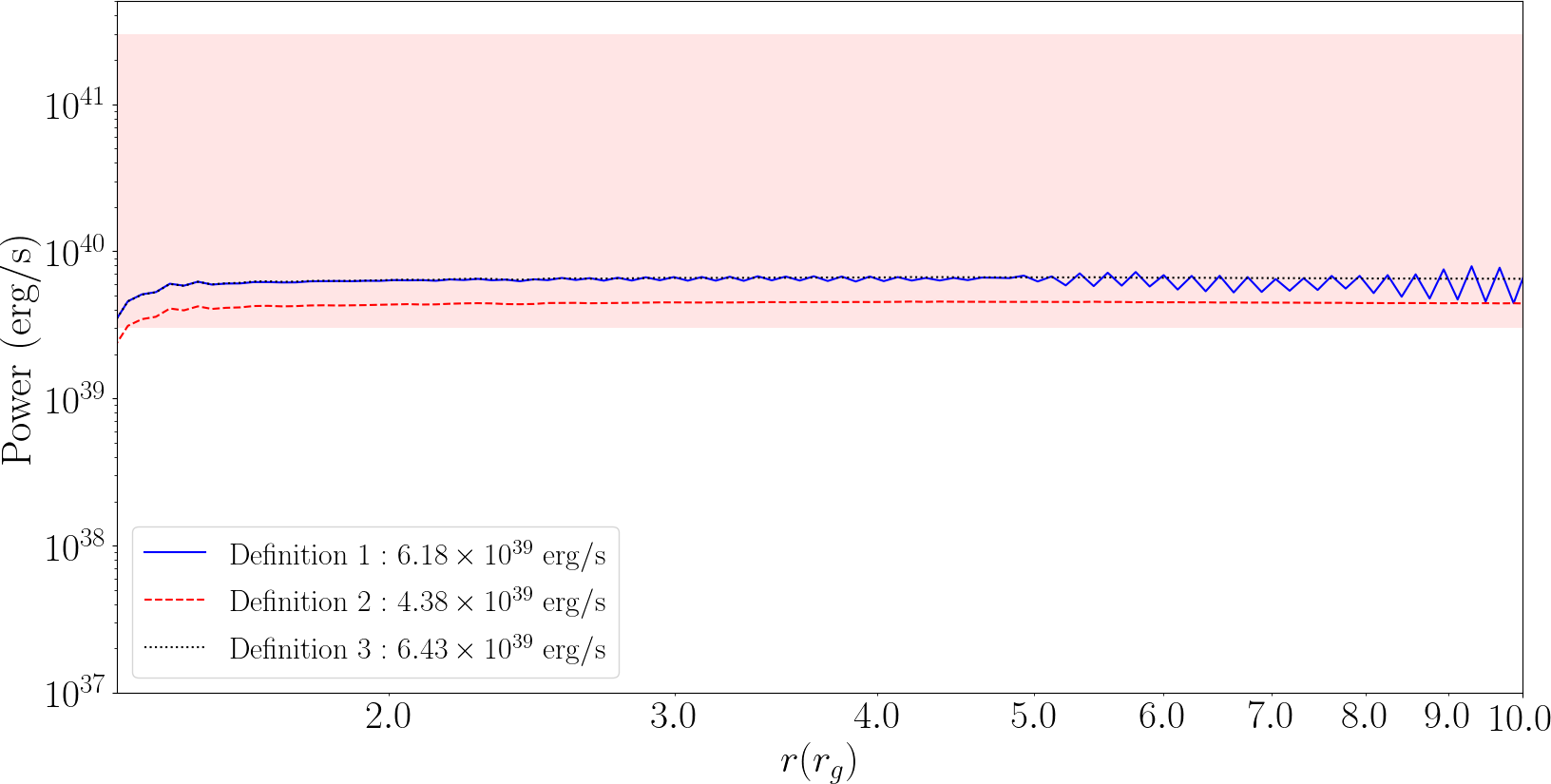}
\caption{SANE}
\end{subfigure}
\begin{subfigure}[b]{0.83\textwidth}
\includegraphics[width=\textwidth]{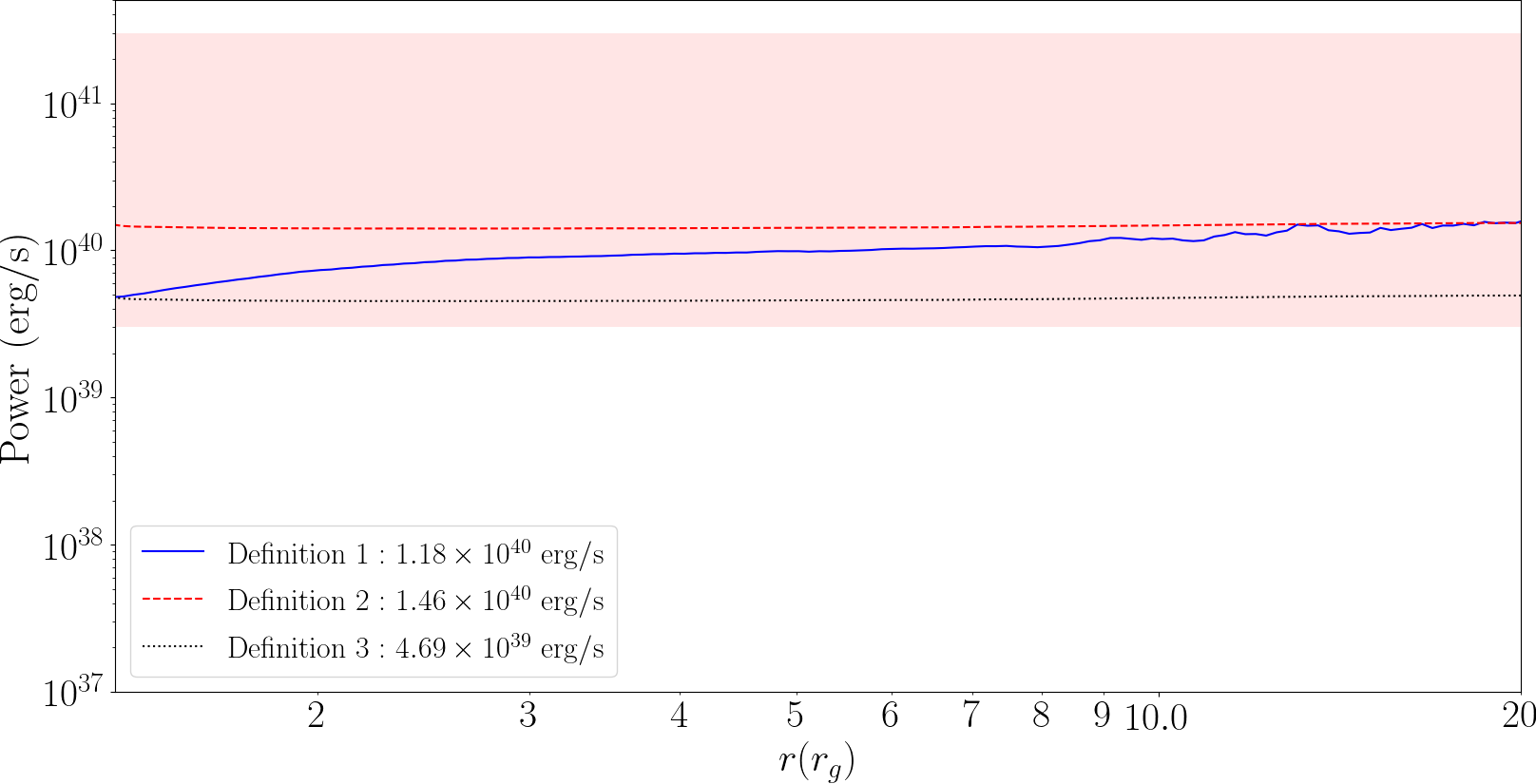}
\caption{MAD}
\end{subfigure}
     \caption{Power profiles for the all definitions of time average. Shaded region indicates ULX luminosity range. Average power for the each definition of power is also mentioned next to the indicated line-type.}
     \label{po}
 \end{figure}

Using eq. (\ref{dpower}) and the above formalism for $r_{eq}$, we have explored the time averaged power profiles for the aforementioned three definitions. Fig. \ref{po} clearly shows that all the power definitions lie within the ULX range for both SANE and MAD simulations. However, the average power is higher for the MAD simulations than SANE. This shows that MAD systems are capable of producing higher outflow power. Nevertheless, the important point to note is that we do not include radiative cooling in the present work, hence the estimated outflow power is only the precursor to ULX luminosities or the upper bound.

 \begin{figure}
\centering
\begin{subfigure}[b]{0.7\textwidth}
\includegraphics[width=\textwidth]{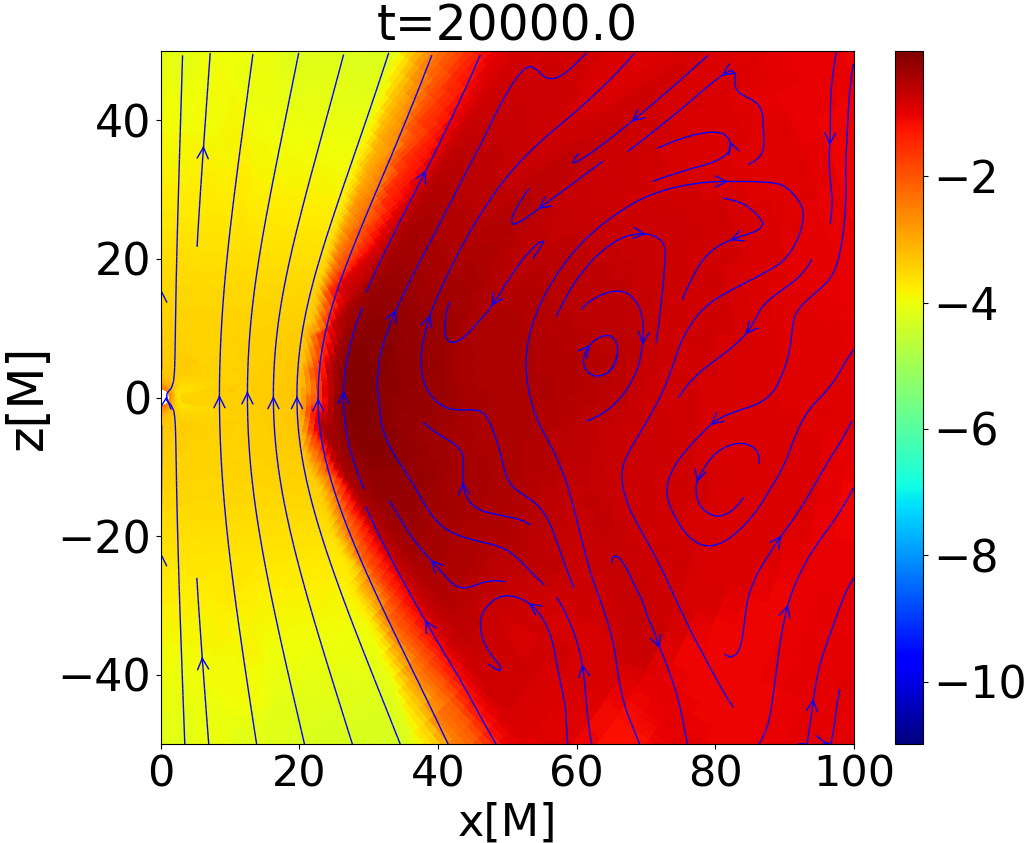}
\caption{}
\end{subfigure}
\begin{subfigure}[b]{0.8\textwidth}
\includegraphics[width=\textwidth]{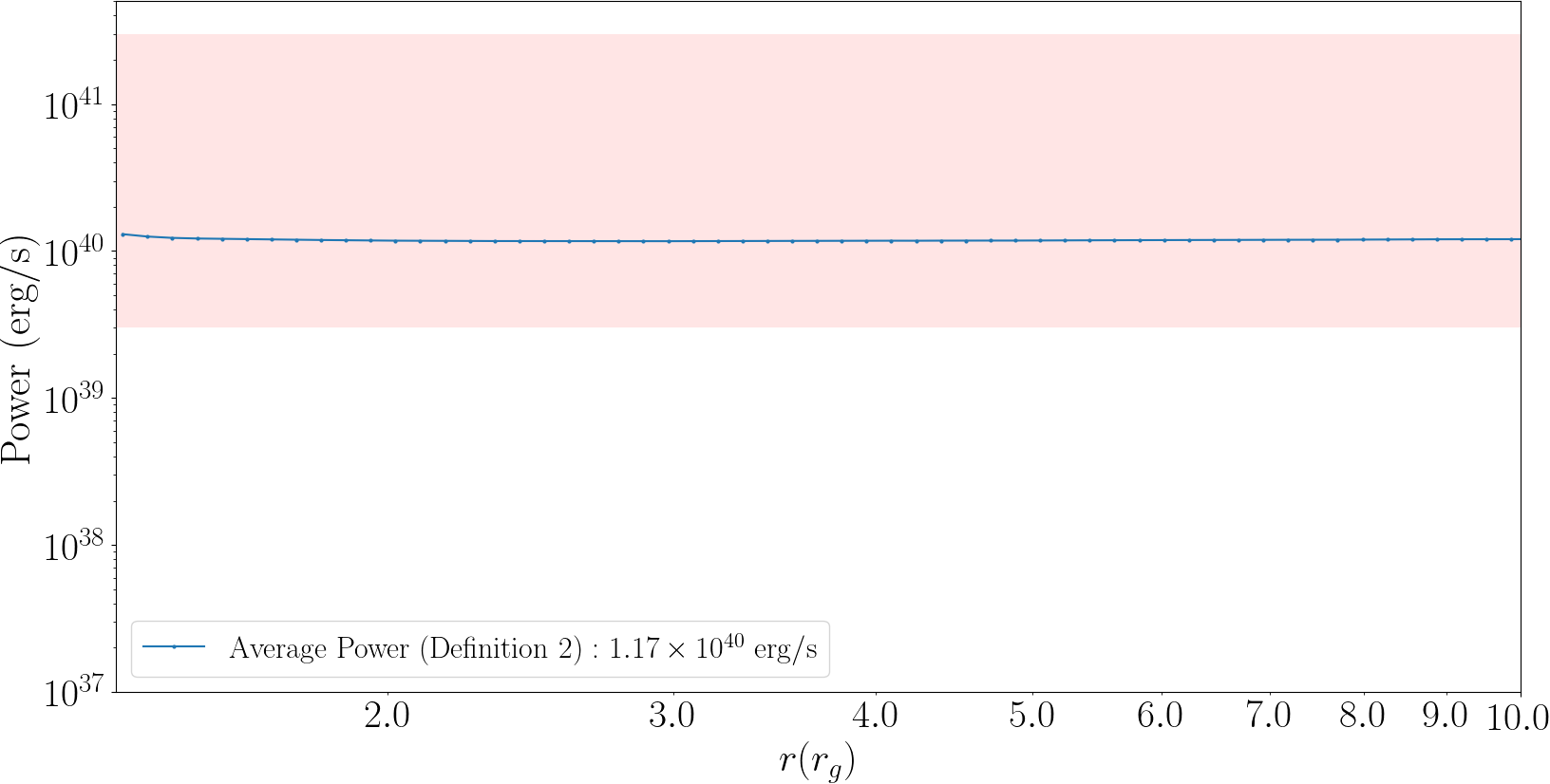}
\caption{}
\end{subfigure}
     \caption{(a) Density contour overplotted with magnetic field streamlines at $t=20000$ $r_g/c$. (b) Outflow power profile according to definition 2.}
     \label{ro}
 \end{figure}

Upon considering the same parameters as Raha et al. \cite{raha} used for their MAD simulation in HARMPI, i.e., the system domain going from 1.29 $r_g$ to 10000 $r_g$ with resolution of $256\times256$, keeping the other MAD simulation parameters same as mentioned above, the obtained density contour and power profile are given by Fig. \ref{ro}. The time averaging for the power profile has been done from 15000 $r_g/c$ to 20000 $r_g/c$, as Raha et al. \cite{raha}. The density contour shows strong poloidal magnetic fields forming a barrier, resulting in a MAD-like configuration. The power profile shows similar outflow power magnitude as obtained in Fig. \ref{po}. This shows that our outflow power remains similar irrespective of computational domain size, resolution and time averaging window, thus exhibiting the robustness of our results. 

 \subsection{Magnetic field profile}

MM19 showed that the fields required to produce ULX luminosities from advective flows are of the order of $10^7$ G. However, to achieve such high fields, they had to consider super-Eddington magnetic fields far away from the black hole.

\begin{figure}
\centering
\includegraphics[width=0.8\textwidth]{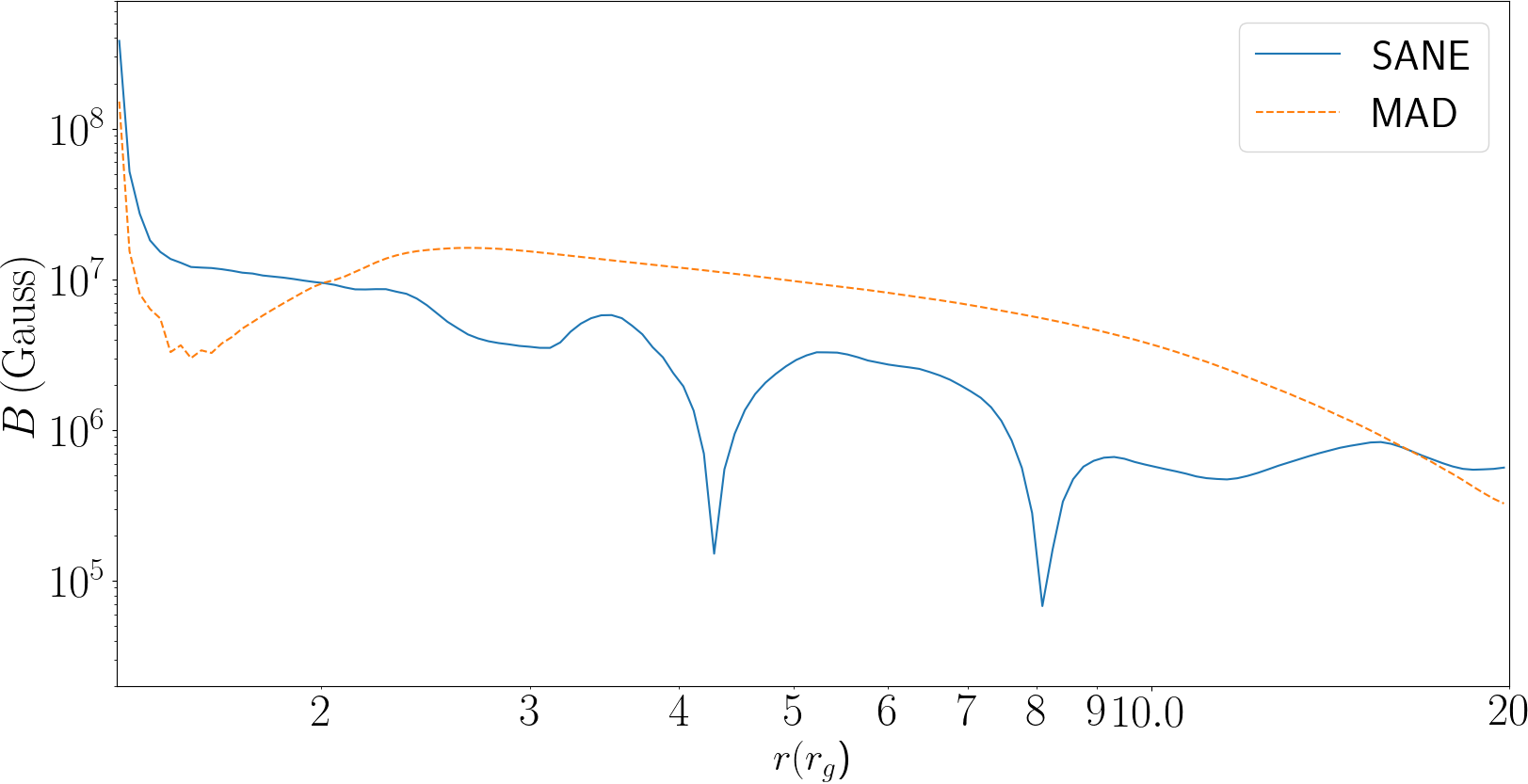}
     \caption{Magnetic field profiles for SANE and MAD simulations.}
     \label{mag}
 \end{figure}

Fig. \ref{mag} shows that the field strength obtained from our simulations is also $\sim$ $10^7$ G, without however any requirement for super-Eddington fields. It is also evident that although the peak magnetic field in the SANE simulations is slightly higher than the MAD simulations very close to the black hole, the higher MAD magnetic fields at other radii lead to higher luminosities compared to SANE. The differences in powers, whatever they be, between SANE and MAD is also reflected in their magnetic fields.

\section{Discussion}
MAD simulations are known to show more than 100\% efficiency. This fact has been used to explain apparent more than 100\% efficient accretion flows in AGNs \cite{rn11}. In our simulations as well, we observe higher outflow power for MAD compared to SANE, even though SANE has higher peak magnetic field. This is because MAD flows are very efficient and are capable of producing strong outflows with lower fields.

\begin{figure}
\begin{subfigure}[b]{0.49\textwidth}
\includegraphics[width=\textwidth]{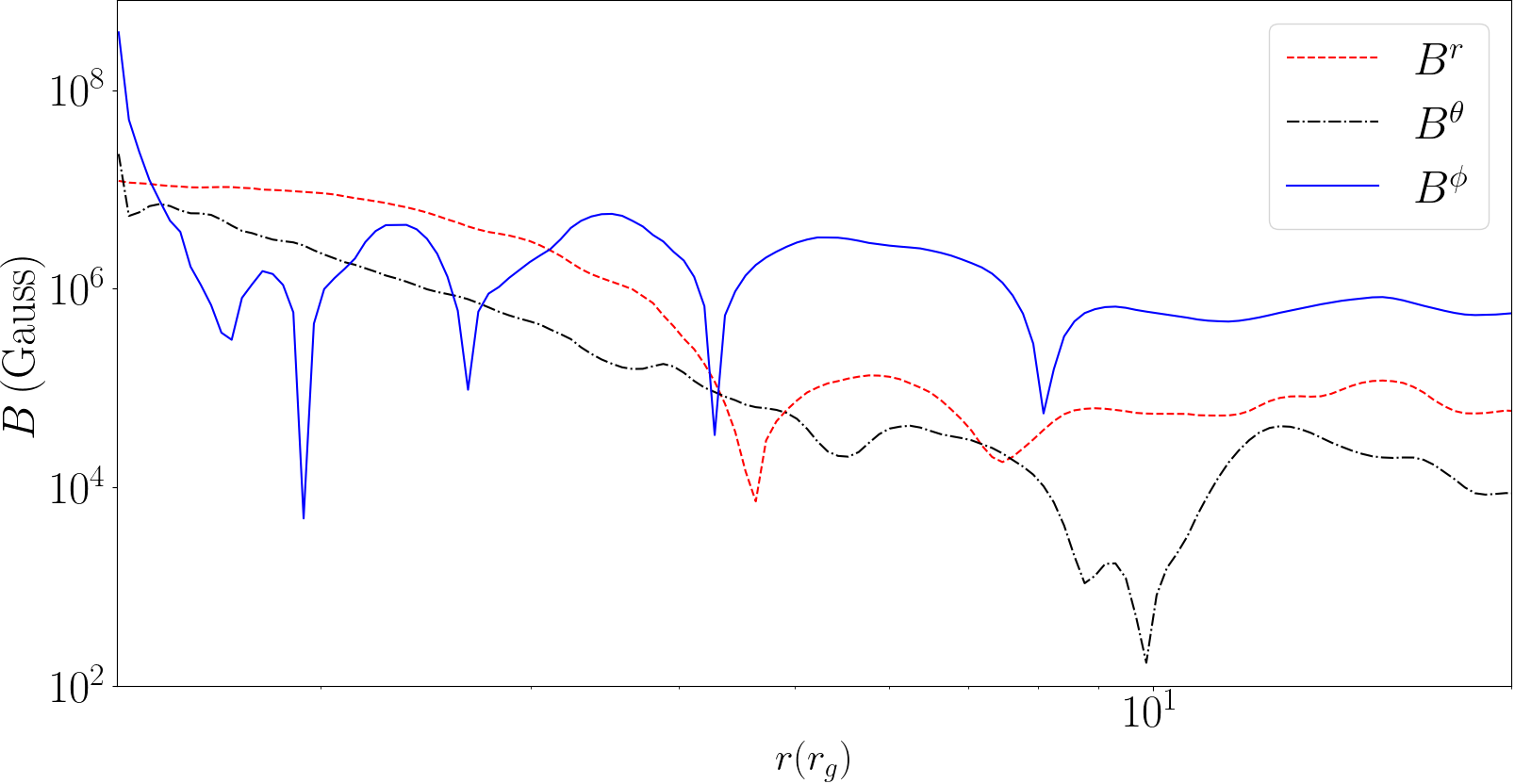}
\caption{SANE}
\end{subfigure}
\begin{subfigure}[b]{0.49\textwidth}
\includegraphics[width=\textwidth]{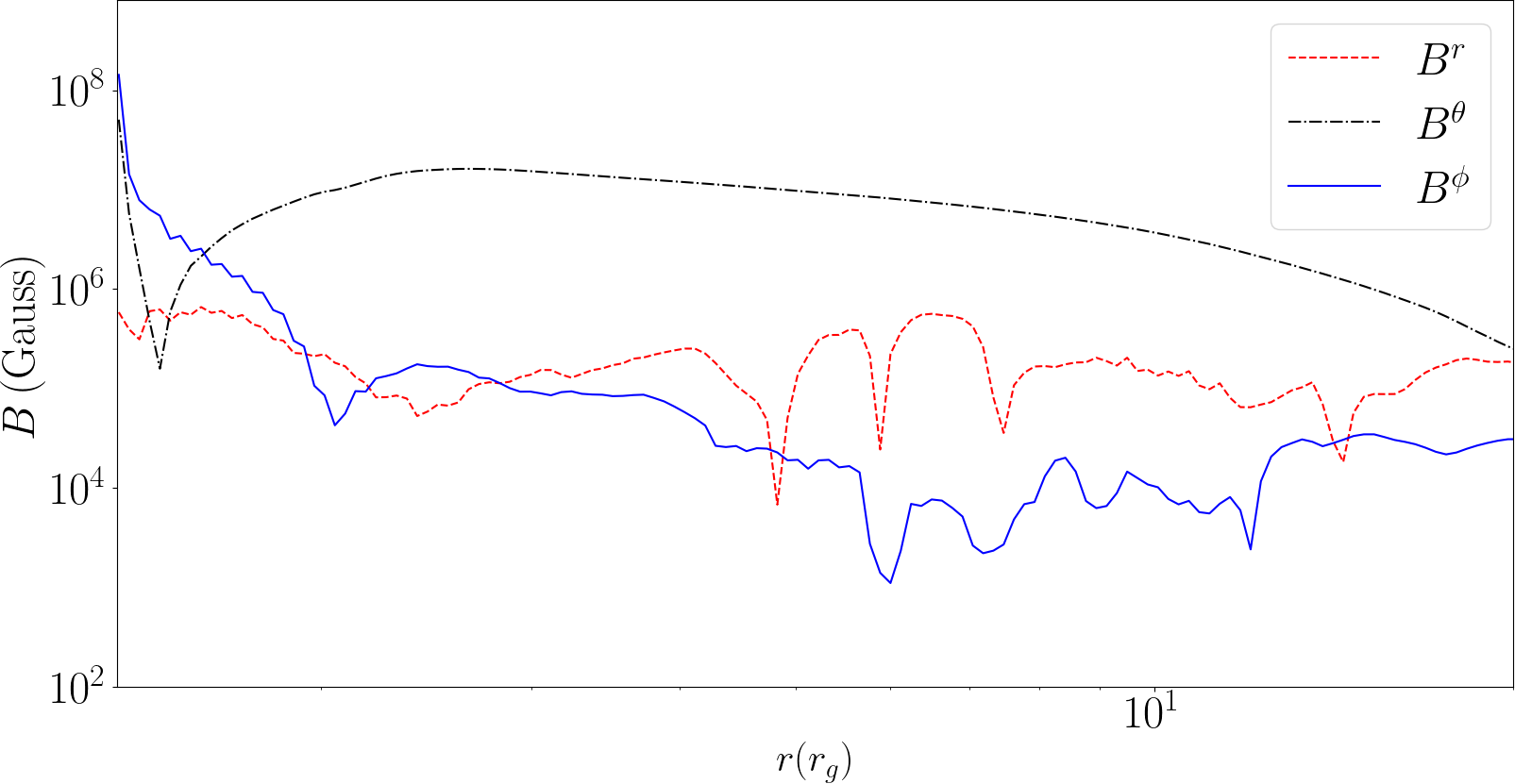}
\caption{MAD}
\end{subfigure}
     \caption{Magnetic field components for SANE and MAD simulations.}
     \label{comp}
 \end{figure}

This result can also be attributed to the fact that magnetic fields evolve differently for MAD and SANE simulations. Fig. \ref{comp} shows the magnetic field components for SANE and MAD simulations. Unlike the SANE system, the MAD system exhibits a significant increase in $B_\theta$ at larger radii which results in sustained magnetic fields in MAD leading to higher outflow power.


\section{Outflow power in the ZAMO frame}

The above analysis is valid for the observers at infinity and the obtained power profiles are flat. Due to this, the driving mechanism of the outflow cannot be attributed to the ergosphere (Blandford-Znajek (BZ) mechanism \cite{bz}), magnetocentrifugal effect of the disk (Blandford-Payne (BP) mechanism \cite{bp}) or simply disk wind. In our study, the combined outflow due to the magnetocentrifugal effect and disk wind is considered to be BP outflow. 

There is no observer-independent way of defining energy of the gravitational field at a point. To study the interaction between the black hole and jet, we need to be in a local frame, which for the Kerr metric is fiducially the zero angular momentum observer (ZAMO) frame. To study the aforementioned outflow mechanisms, we investigate outflow power in the ZAMO frame. 

\subsection{Formalism}

The outward energy flux is defined as:
\begin{equation}
    \Dot{E}(r)=-\int\sqrt{-g}T^\mu_\nu u^\nu e_\mu d\theta d\phi,
\end{equation}
where $u^\nu$ and $e_\mu$ are the fluid four velocity and local four unit vector respectively, $T^\mu_\nu$ is same as in eq. (4).

The jet region is considered to be the part of the simulation domain in which $-T^\mu_\nu>0$ \cite{rp}. We only consider outflow in the radial direction, which leads to $e_\mu=(0,1/g^{rr},0,0)$, where $g^{rr}$ is the `$rr$' component of the contravariant Kerr metric.
We have run simulations for various values of black hole spin parameter, namely, $a=0.9,0.8,0.6,0.0$, to study the effect of $a$ on the outflow power.

\subsection{Power profiles}
\begin{figure}
\centering
\includegraphics[width=0.9\textwidth]{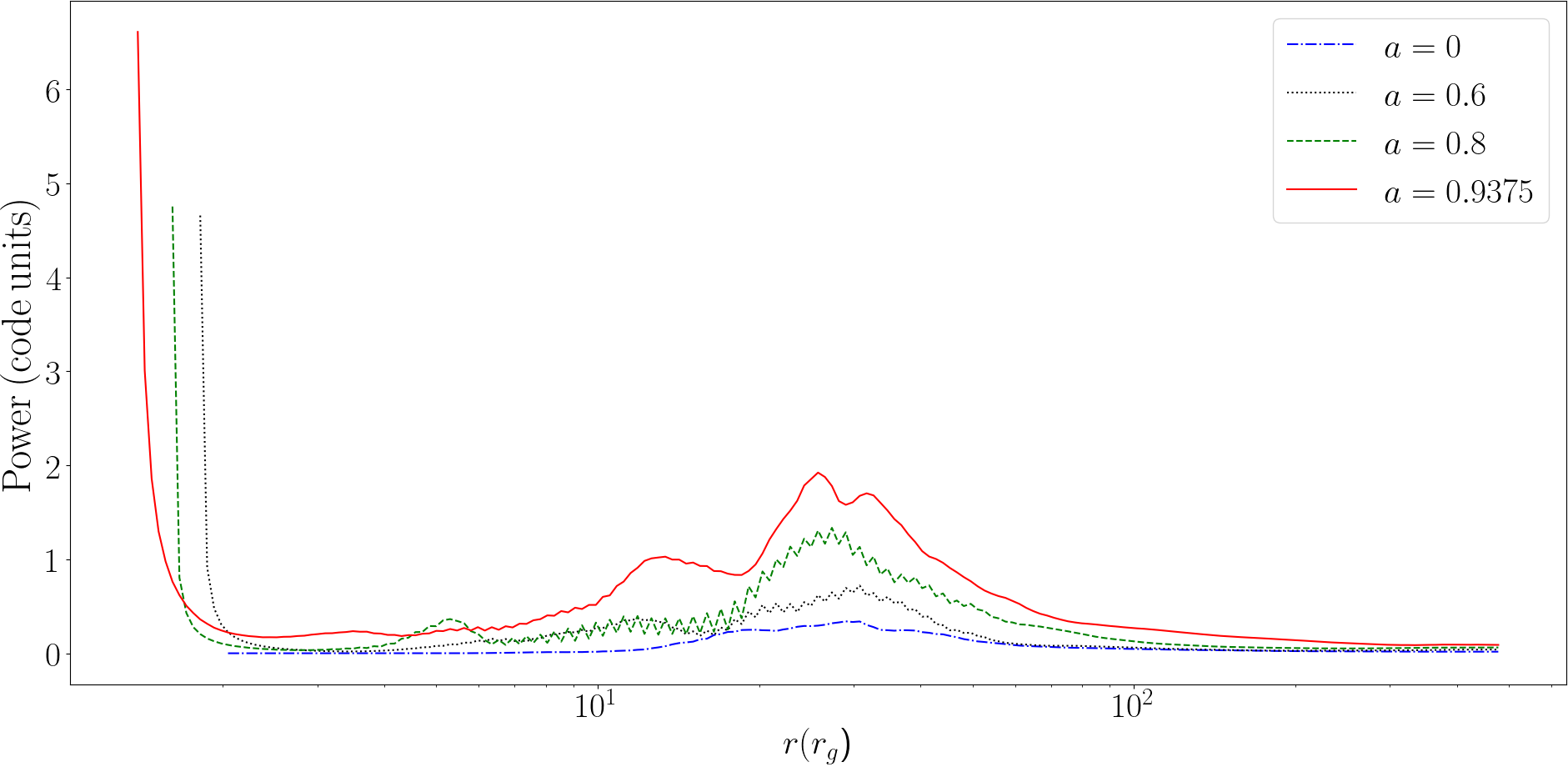}
     \caption{Power profiles for ZAMO observer for different $a$ values.}
     \label{zamo}
\end{figure}

Fig. \ref{zamo} shows the obtained power profiles for the different $a$ values. It is clearly evident that the profile has two contributing components, a peak near the horizon that dies down by the end of the ergosphere and another smaller component around $r=30r_g$.

The peak near the horizon decreases monotonically with $a$ and reaches a minimum at $r=2r_g$, i.e. the ergosphere. This can be interpreted as the Blandford-Znajek mechanism, which states that jets/outflows are powered by the spin of the black hole due to the strong inertial frame dragging effect. Particles entering the ergosphere tap the energy of the angular momentum reservoir of the black hole and are kicked out along the magnetic field lines. The field lines themselves are twisted due to the frame dragging which introduces magnetic tension force along the axis of rotation of the black hole. The magnetic tension force collimates the jet outflow.

This also explains the peak value decreasing with $a$. For slowly rotating black holes, there is not enough energy to tap from the black hole reservoir. This confirms that for non-rotating black holes, there will not be any emission due to the BZ mechanism. This is also evident from Fig. \ref{zamo}.

The second peak near $r=30r_g$ is attributed to the BP process. This is because BP states that in the presence of strong magnetic fields, matter in the accretion disk starts to move along the magnetic field lines under the effect of magnetocentrifugal forces. Due to the acceleration of the particles, the field line can only support the particle flow till the Alfv\'en radius, after  the magnetic tension force dominates over the magnetic pressure force and collimates the flow. Thus BP mechanism is valid for emissions from any part of the disk and is not strictly dependent on the central compact object. 

However since the particles are not taping any energy reservoir in BP mechanism, the outflow power is lower than that of the BZ process. BZ on the other hand is an active process, in which the black hole acts as the initial energy source leading to stronger outflows.

\section{Conclusion}

Our simulations show that highly magnetised advective accretion flows can indeed produce high outflow power, well within the observed ULX luminosity range. This shows that the peculiar properties of hard state ULXs can be explained without invoking principles like modifying the Eddington luminosity by changing the electron scattering cross-section or considering ULXs to be intermediate mass black holes, only very few of which are currently known. 

ZAMO frame results show that studying accretion-outflows by changing observer frames might give insights into the outflow generation mechanisms from accreting sources. Our results show two distinct peaks in the power profiles which can be attributed to BZ and BP mechanisms for outflow generation.

\end{document}